\documentclass{llncs}

\usepackage{xspace,amsmath,url}

\newcommand{\ceil}[1]{\lceil #1 \rceil}
\newcommand{\floor}[1]{\lfloor #1 \rfloor}

\def \bxor{\textsf{xor}\xspace}
\def \band{\textsf{and}\xspace}

\newcommand{\uxor}{\mathrel{^\wedge}}
\newcommand{\uand}{\mathrel{\&}}
\newcommand{\uor}{\mathrel{|}}
\newcommand{\unot}{\mathop{\sim}}
\newcommand{\fsize}{f}
\newcommand{\bsize}{b}
\newcommand{\len}{r}
\newcommand{\word}[1]{#1}
\newcommand{\sabound}{$O((n \log\sigma/\log n)\ceil{m \log (k +  \log n / \log\sigma) / w} + n^{\varepsilon})$\xspace}
\newcommand{\acbound}{$O(\frac{n}{\floor{w/(m\log\sigma)}} (1 + \log \min(k,\sigma) \log m / \log\sigma))$\xspace}
\newcommand{\wrbound}{$O(\frac{n}{\floor{w/(m\log\sigma)}}\log \min(m, \log w / \log\sigma))$\xspace}
\newcommand{\fword}[1]{$(#1)$-word}

\begin{document}

\title{Approximate pattern matching with $k$-mismatches in packed text\thanks{This paper is an extended version of the article that appeared in \emph{Information Processing Letters} 113(19-21):693-697 (2013), http://dx.doi.org/10.1016/j.ipl.2013.07.002}}

\author{Emanuele Giaquinta\inst{1} \and Szymon Grabowski\inst{2} \and Kimmo Fredriksson\inst{3}}

\institute{Department of Computer Science, University of Helsinki, Finland
	  \email{emanuele.giaquinta@cs.helsinki.fi} \and
	  Lodz University of Technology, Institute of Applied Computer Science, Al.\ Politechniki 11, 90--924 {\L}\'od\'z, Poland
	  \email{sgrabow@kis.p.lodz.pl} \and
          School of Computing, University of Eastern Finland, P.O. Box\ 1627, FI-70211 Kuopio, Finland
          \email{kimmo.fredriksson@uef.fi}
}

\maketitle

\begin{abstract}
Given strings $P$ of length $m$ and $T$ of length $n$ over an alphabet
of size $\sigma$, the string matching with $k$-mismatches problem is
to find the positions of all the substrings in $T$ that are at Hamming
distance at most $k$ from $P$. If $T$ can be read only one character
at the time the best known bounds are $O(n\sqrt{k\log k})$ and $O(n +
n\sqrt{k/w}\log k)$ in the word-RAM model with word length $w$.
In the RAM models (including $AC^0$ and word-RAM)
it is possible to read up
to $\floor{w / \log \sigma}$ characters in constant time if the
characters of $T$ are encoded using $\ceil{\log \sigma}$ bits. The
only solution for $k$-mismatches in packed text works in \sabound
time, for any $\varepsilon > 0$. We present an algorithm that runs in
time \acbound in the $AC^0$ model if $m=O(w / \log\sigma)$ and $T$ is
given packed. We also describe a simpler variant that runs in time
\wrbound in the word-RAM model. The algorithms improve the existing
bound for $w = \Omega(\log^{1+\epsilon}n)$, for any $\epsilon > 0$. Based on the introduced technique, we present
algorithms for several other approximate matching problems.
\end{abstract}

\section{Introduction}
\noindent
The string matching problem consists in reporting all the occurrences
of a pattern $P$ of length $m$ in a text $T$ of length $n$, both strings over a common alphabet. The
occurrences may be exact or approximate according to a specified
matching model.
For most matching problems, all the characters from the text and the
pattern need to be read at least once in the worst case; hence, if
they are read one at a time, the worst-case lower bound is
$\Omega(n)$.
Interestingly, for some standard problems (e.g., exact pattern
matching) it is possible to achieve a sublinear search time, for short
patterns, even in the worst case, if the word-RAM computational model
is assumed and the text is {\em packed}. In a packed encoding, the
characters of a string are stored adjacently in memory and each
character is encoded using $\log\sigma$
bits\footnote{Throughout the paper, all logarithms are in base 2.
  W.l.o.g. we also assume that $\sigma$ is a power of two.}, where
$\sigma$ is the alphabet size. A single machine word, of size $w \geq
\log n$ bits, thus contains up to $\alpha = \floor{w / \log\sigma}$
characters.
While the word size of current architectures is $64$ (which, for
example, permits up to $32$ DNA symbols to be encoded in a word), there are also
vector instruction sets where the word size is larger, such as SSE and
AVX ($128$ and $256$ bits) or the Intel Xeon Phi coprocessor ($512$
bits).

For this setting and the exact string matching problem, several
sublinear-time algorithms have been given in recent
years~\cite{Fredriksson2002,Bille2011,Belazzougui2012jda,Ben-KikiBBGGW2011,BreslauerGG2012}.

In this paper we study the string matching with $k$-mismatches problem
in the packed scenario. This problem is to find
the positions of all the substrings in $T$ that are at Hamming
distance at most $k$ from $P$, i.e., that match $P$ with at most $k$
mismatches. For this problem, the best known bounds in the worst-case
are $O(n\sqrt{k\log k})$ time for the algorithm by Amir et
al.~\cite{DBLP:journals/jal/AmirLP04} and $O(n + n\sqrt{k/w}\log k)$
time for its implementation based on word-level
parallelism~\cite{DBLP:journals/ejc/FredrikssonG13}. One classical
result in the word-RAM model that is also practical is Shift-Add~\cite{DBLP:journals/cacm/Baeza-YatesG92}.
The best worst-case bound of this algorithm, based on the Matryoshka counters
technique~\cite{DBLP:journals/ipl/GrabowskiF08}, is $O(n\ceil{m/w})$.

In~\cite{Fredriksson2002} Fredriksson presented a Shift-Add variant,
based on the super-alphabet technique, that works in \sabound time,
for any $\varepsilon > 0$. To our knowledge, this is the only solution
for the $k$-mismatches problem that works on packed text and that
achieves sublinear time complexity when $m$ and $k$ are sufficiently
small.

In this work, we present an algorithm for the $k$-mismatches problem
that runs in time \acbound in the $AC^0$ model for $m\le \alpha$ if
$T$ is given packed. We also describe a simpler variant that runs in
time \wrbound in the word-RAM model. In particular, it achieves
sublinear worst-case time when $m\log\sigma\log\min(m, \log w / \log\sigma)
= o(w)$. Note that for $w = \Theta(\log n)$ Fredriksson's solution is
better, but our algorithm dominates if $w = \Omega(\log^{1+\epsilon}n)$, for any $\epsilon > 0$, or, more precisely, if $w=\omega(\log n \log \log w)$.

\section{Basic notions and definitions}
\noindent
Let $\Sigma = \{0, 1, \ldots, \sigma - 1\}$ denote an integer alphabet
and $\Sigma^m$ the set of all possible
sequences of length $m$ over $\Sigma$.
$S[i], i \geq 0$, denotes the $(i+1)$-th character of string $S$, and
$S[i\,\ldots\,j]$ its substring between the $(i+1)$-st and the
$(j+1)$-st characters (inclusive).

The $k$-mismatches problem consists in, given a pattern (string) $P$
of length $m$ and a text (string) $T$ of length $n$, reporting all the
positions $0 \leq j \leq n-m$ such that $|\{0 \leq h < m\ :\ T[j +
  h]\neq P[h]\}|\le k$, i.e., such that the Hamming distance between
$P$ and the substring $T[j\,\ldots\,j+m-1]$
is at most $k$.

The word-RAM model is assumed, with machine word size $w \ge \log
n$. We use some bitwise operations following the standard notation as
in C language: $\&$, $|$, $^\wedge$, $\sim$, $<<$, $>>$ for
\texttt{and}, \texttt{or}, \texttt{xor}, \texttt{not}, \texttt{left
  shift} and \texttt{right shift}, respectively.

\section{Operations on words}

We define a \fword{f} as a machine word logically divided into
$\floor{w/f}$ fields of $\fsize$ bits. Given a \fword{f} $\word{W}$,
we denote with $\word{W}[i]$ its $i$-th field, for
$i=1,\ldots,\floor{w/f}$. The most significant bit in a field is
called the \emph{top} bit.
A field where only the top bit is set is thus equal to $2^{\fsize-1}$.
We also define, for a given field size
$\fsize$, the \emph{mask} $V_{\fsize}$ where $\word{V}_{\fsize}[i]
= 2^{\fsize-1}$, for $i=1,\ldots,\floor{w/f}$. We define the following
primitives on words:

\medskip
\noindent\emph{find non-zero fields} (\textsf{fnf}($\word{A},\fsize$)): given a
\fword{\fsize} $\word{A}$, return a \fword{\fsize} $\word{A'}$
such that $\word{A'}[i]$ is equal to $2^{\fsize - 1}$ if
$\word{A}[i]\neq 0$, and to $0$ otherwise, for
$i=1,\ldots,\floor{w/\fsize}$.
\smallskip

\noindent How to implement this primitive in $O(1)$ time was presented
in~\cite[Sect.~4]{BreslauerGG2012}, but we give a simpler method, in
three simple steps and in constant time.
\begin{enumerate}
\item $\word{W}\leftarrow \word{A} \uand \unot\word{V_{\fsize}}$
\item $\word{X}\leftarrow \word{V_{\fsize}} - \word{W}$
\item $\word{A'}\leftarrow (\unot\word{X} \uor \word{A}) \uand \word{V_{\fsize}}$
\end{enumerate}
The first two steps generate a \fword{\log\sigma} $\word{X}$ such
that $\word{X}[i] = 2^{\log\sigma-1}$ if $\word{A}[i]$ with the
top-bit masked is zero, and $ < 2^{\log\sigma-1}$ otherwise. It is
then not hard to see that $\word{A}[i]$ is non-zero iff either
$\word{A}[i]\ge 2^{\log\sigma-1}$ or $(\unot\word{X})[i]\ge
2^{\log\sigma-1}$, from which follows the correctness of the last
step.

\medskip
\noindent\emph{sideways addition} (\textsf{sa}($\word{A}$)): given a word $\word{A}$,
return the number of bits set in $\word{A}$.
\smallskip

\noindent This primitive is a well-known bitwise operation, also known
as popcount. The folklore
method~\cite{DBLP:conf/wea/Vigna08}
to compute it in the word-RAM model has $O(\log \log w)$ time complexity.

\medskip
\noindent\emph{interleaved blockwise sideways addition} (\textsf{ibsa}($A,\fsize,\bsize$)):
Given a \fword{\fsize} $A$, such that $\fsize$ divides $\log\sigma$ and
only the top bit of each field may be set, and a power of two $b$,
return a \fword{\fsize} $A'$ such that $\word{A'}[i]$ is equal to
$$
\sum_{j=0}^{\min(b,\ceil{i/(\log\sigma / f)})-1} X[i-j(\log\sigma / f)]\,,
$$
where $X = A>> (\fsize - 1)$, i.e., the $i$-th field contains the
number of bits set in the sequence of $\min(b,\ceil{i/(\log\sigma / f)})$
fields in $\word{A}$ spaced by $\log\sigma$ bits ending at $i$.
\smallskip

\noindent This operation is a variant of the parallel prefix-sum operation
described in~\cite{HS1986} and can be implemented in $O(\log b)$
steps, where the $j$-th step computes
$$
A'_j = \begin{cases}
  A'_{j-1} + (A'_{j-1} << (2^{j-1}\times \log\sigma)) & \text{if } j > 0 \\
  A >> (\fsize-1) & \text{otherwise} \\
\end{cases}
$$
Since $\fsize \ge \log(\bsize + 1)$ does not necessarily hold, the top
bits of all the fields are masked out before each addition and
restored afterwards. In this way, if a sum is $\ge 2^{\fsize-1}$, its
encoded value is $\ge 2^{\fsize-1}$ but the exact value is
undetermined.

\medskip
\noindent\emph{blockwise sideways addition} (\textsf{bsa}($\word{A},
\fsize, \bsize)$): Given a \fword{\fsize} $\word{A}$, such that only
the top bit of each field may be set, and a power of two $\bsize$,
return a \fword{b\fsize} $\word{A'}$ such that $\word{A'}[i]$ is equal to
$$
\sum_{j=0}^{\bsize-1} X[i\bsize - j]\,,
$$
where $X = A>> (\fsize - 1)$, i.e., the
$i$-th field contains the number of bits set in the block of
$\bsize$ fields in $\word{A}$ ending at $i\bsize$.
\smallskip

\noindent This operation can be implemented in time $O(\log
\min(b,\log w / \fsize))$ in word-RAM and in time $O(\log b)$ in
$AC^0$ using the following method. We assume that the word size $w$ is
a power of two.
Let $\len$ be the smallest power of two greater than or equal to
$\log (w + 1) / \fsize$. The first step consists in computing a word logically
divided into fields of $\min(\bsize, \len)\fsize$ bits, such that each
field contains the number of bits set in the corresponding
$\min(\bsize, \len)$ fields in the original word. This widening
operation can be performed in $\log \min(\bsize, \len) = O(\log
\min(b,\log w / \fsize))$ steps using simple bitwise operations and $\log \min(\bsize, \len)$ masks.

Since both $\len$ and $\bsize$ are a power of two, each block of $\bsize\fsize$ bits spans an
integral number of fields of $\min(\bsize, \len)\fsize$ bits. Observe
that there can be at most $w$ bits set in a word, so $\len\fsize$ bits
are enough to encode the total number of bits. If $\bsize \le \len$,
then since $\bsize$ is a power of two after the last widening step we
have a word divided into fields of $\bsize\fsize$ bits, each one
containing the desired number of ones. Otherwise, if $\bsize > \len$,
we compute the prefix sum of the sequence of numbers given by the
fields, i.e., we store into each field the sum of the previous fields
including itself. In word-RAM we do this by performing a
multiplication (which is $O(1)$), with the mask $0^{\len\fsize
  -1}1\ldots0^{\len\fsize -1}1$. Instead, in $AC^0$ we use again the
parallel prefix-sum algorithm described in~\cite{HS1986}, which is
$O(\log b)$. It is not hard to see that, after this operation, the
number of bits in a block is equal to the last field of the block
minus the last field of the previous block. This operation can be
implemented in parallel for all the blocks with a shift and a subtraction.
Finally, to obtain the desired output word we reset to zero all the
fields but the last of each block, using an \band with the mask
$1^{\len\fsize}0^{(\bsize - \len)\fsize}\ldots 1^{\len\fsize}0^{(\bsize -  \len)\fsize}$,
and shift the word to the right by $(\bsize - \len)\fsize$ bits.

The pseudocode of \textsf{bsa} in word-RAM is the following:
\begin{enumerate}
\item $X\leftarrow A>> (\fsize - 1)$
\item $l\leftarrow \fsize$
\item \textbf{for} $i\leftarrow 1$ \textbf{to} $\log \min(b,\len)$ \textbf{do}
\item \qquad $H\leftarrow X>> l$
\item \qquad $X\leftarrow (X\uand 0^l1^l\ldots 0^l1^l) + (H\uand 0^l1^l\ldots 0^l1^l)$
\item \qquad $l\leftarrow l \times 2$
\item \textbf{if} $\min(\bsize,\len) = \len$ \textbf{then}
\item \qquad $X\leftarrow X \times 0^{\len\fsize-1}1\ldots0^{\len\fsize -1}1$
\item \qquad $X\leftarrow X - (X<< \bsize\fsize)$
\item \qquad $A'\leftarrow (X\uand 1^{\len\fsize}0^{(\bsize - \len)\fsize}\ldots 1^{\len\fsize}0^{(\bsize -\len)\fsize})>> (\bsize - \len)\fsize$
\end{enumerate}


\medskip
\noindent\emph{parallel minima (maxima)~\cite{PS1980}} (\textsf{pmin
  (pmax)}($\word{A},\word{B},f$)): Given two \fword{f}s $\word{A}$ and $\word{B}$, return a
\fword{f} $\word{W}$ such that $\word{W}[i]$ is equal to $2^{f-1}$
if $\word{A}[i]\le \word{B}[i]$ ($\word{A}[i]\ge \word{B}[i]$), and to
$0$ otherwise, for $i=1,\ldots,\floor{w/f}$. \textsf{pvmin}
(\textsf{pvmax}) is similar, but $\word{W}[i]$ is equal to
$\min(\word{A}[i],\word{B}[i])$ ($\max(\word{A}[i],\word{B}[i])$).
\smallskip

\noindent These operations can be implemented in constant time, as demonstrated by the following code
(\textsf{pmin}):

\begin{enumerate}
\item $\word{T_A} \leftarrow \word{A} \uand \word{V_f}$
\item $\word{T_B} \leftarrow \word{B} \uand \word{V_f}$
\item $\word{A'} \leftarrow \word{A} \uand \unot \word{V_f}$
\item $\word{A''} \leftarrow (\word{B} \uor \word{V_f}) - \word{A'}$
\item $\word{H_1} \leftarrow \unot\word{T_A} \uand \word{T_B}$
\item $\word{H_2} \leftarrow \word{A''} \uand (\word{T_A} \uxor \word{T_B} \uxor \word{V_f})$
\item $\word{W} \leftarrow (\word{H_1} \uor \word{H_2}) \uand \word{V_f}$
\end{enumerate}

All the given bounds do not include the time to compute the used masks, if any.

\section{The algorithm}
\label{sec:the_algorithm}

\noindent
We start the presentation with a simple idea, which is then extended
and modified in some ways. Consider two
\fword{\log\sigma}s $\word{A}$ and $\word{B}$, each containing a
packed string of length $m\le \alpha$ in its $m\log\sigma$ least significant
bits (i.e., each field of $\log\sigma$ bits encodes a character). The
higher bits in both words, if any, are all 0s
We perform the \bxor operation of $\word{A}$ and $\word{B}$ and the
number of non-zero fields in the result is exactly the Hamming
distance between the two strings. To count the number of such fields,
we first convert, using the \textsf{fnf} operation, each non-zero
field into a field with only the top-bit set, and then count the number
of bits set using the \textsf{sa} operation.
The procedure to compute the Hamming distance of $\word{A}$ and
$\word{B}$ can thus be implemented in time $O(\log\log w)$ with the
following operations:
\begin{enumerate}
\item $\word{X}\leftarrow \word{A} \uxor \word{B}$
\item $\word{A'}\leftarrow \textsf{fnf}(\word{X}, \log\sigma)$
\item \textbf{return} $\textsf{sa}(\word{A'})$
\end{enumerate}
For arbitrary $m$, observe that the packed encoding of a string of
length $m$ requires $\ceil{m\log\sigma / w}$ words, and the Hamming
distance between two such strings can be computed by running the above
procedure for each word and summing the outputs.

Using this method, we can obtain an algorithm for the string matching
with $k$-mismatches problem that runs in $O(n \ceil{m\log \sigma /
  w}\log\log w)$ time for any $m$. Note that the resulting algorithm
is also practical and compares favorably with the classical Shift-Add~\cite{DBLP:journals/cacm/Baeza-YatesG92}
for small alphabets
and large $k$, although it is less flexible (no support for classes of
characters). It is also worth noting that recent processors include a
\textsc{POPCNT} instruction to compute the sideways addition of a
word, so the $\log\log w$ term disappears in practice.

We now show how to apply the described ideas in an (improved)
algorithm for the $k$-mismatches problem on packed text for short
patterns. In the following, we shall assume $m\le \alpha$. Our method
exploits a general technique~\cite{DBLP:journals/jea/HyyroFN05} to
increase the parallelism in string matching algorithms based on
word-level parallelism. We present a solution in the $AC^0$
model and a simpler variant in the word-RAM model.
We start with the word-RAM algorithm.
Let $\bar{m}$ be the smallest power of two greater than
or equal to $m$ and let $\ell = \floor{w/(\bar{m}\log\sigma)}$. We
first preprocess the pattern $P$ to create a word $\word{A}$ with
$\ell$ copies of $P$ of length $\bar{m}\log\sigma$ starting from the
least significant bit. The last $\bar{m}-m$ fields of each copy are
set to zero. We perform this padding because the \textsf{bsa} and
\textsf{ibsa} operations which we shall use require the size of the
blocks to be a power of two. Let
$\word{B}_i$ be the word containing the packed encoding of the
substrings $T[j+s\bar{m}\,\ldots\,j+s\bar{m}+m-1]$, for
$s=0,\ldots,\ell-1$, where $j = \ell\floor{i/m}m + i\bmod m$, with
$\bar{m}-m$ zero (padding) fields every $m$ fields (i.e., at the end
of each substring). For example, if
$\sigma = 4$, $m=3$, $T=01\,10\,11\,11\,10\,11\,01\,10$ and $w=16$, then we have
$B_0 = 01\,10\,11\,\#\,10\,11\,01\,\#$ and $B_1 =
10\,11\,11\,\#\,11\,01\,10\,\#$, where $\#$ denotes a padding field.
Note that because of this partitioning we do not process all the text
substrings in linear order. The word $B_i$ can be computed in constant
time by extracting the substring $T[j\,\ldots\,j+\ell\bar{m}-1]$ from
the packed text and clearing the padding fields with a mask.
Our search algorithm performs the following main steps, for each $0\le i < n / \ell$:
\begin{enumerate}
\item $\word{X}\leftarrow \word{A} \uxor \word{B}_i$
\item $\word{A'}\leftarrow \textsf{fnf}(\word{X}, \log\sigma)$
\item $M\leftarrow \textsf{pmin}(\textsf{bsa}(\word{A'}, \log\sigma, \bar{m}), K, \fsize)$
\item $\textbf{report}(M)$
\end{enumerate}
where $\fsize = \bar{m}\log\sigma$ and $K$ is a
\fword{\fsize} with a copy of the integer $k$ in each field.
At each iteration, our algorithm processes $\ell$
substrings of $T$ in parallel using the technique to compute the
Hamming distance of two words described before. First, we perform the
\bxor and \textsf{fnf} operations to identify the mismatches for the
$\ell$ substrings encoded in $B_i$.
\begin{figure}[!t]
\begin{center}
\begin{scriptsize}
\renewcommand{\tabcolsep}{0.09cm}
\begin{tabular}{|rll|}
\hline
$A$ & $\leftarrow$ & $01\,10\,01\,\underline{00}\,01\,10\,01\,\underline{00}$ \\
$B_0$ & $\leftarrow$ & $01\,10\,11\,\underline{00}\,10\,11\,01\,\underline{00}$ \\
$X\leftarrow A\uxor B_0$ & $=$ & $00\,00\,10\,\underline{00}\,11\,01\,00\,\underline{00}$ \\
\hline
$A'\leftarrow$ \textsf{fnf}($X, 2$) & & \\
\hline
$A$ & $\leftarrow$ & $00\,00\,10\,\underline{00}\,11\,01\,00\,\underline{00}$ \\
$V$ & $\leftarrow$ & $10\,10\,10\,\underline{10}\,10\,10\,10\,\underline{10}$ \\
$W\leftarrow A \uand \unot V$  & $=$ & $00\,00\,00\,\underline{00}\,01\,01\,00\,\underline{00}$ \\
$X\leftarrow V - W$ & $=$ & $10\,10\,10\,\underline{10}\,01\,01\,10\,\underline{10}$ \\
$A'\leftarrow (\unot X \uor A) \uand V$ & $=$ & $00\,00\,10\,\underline{00}\,10\,10\,00\,\underline{00}$ \\
\hline
$X\leftarrow \textsf{bsa}(\word{A'}, 2, 4)$ & $=$ & $00000001\,\,\,\,\,00000010$ \\
$K$ & $\leftarrow$ & $00000001\,\,\,\,\,00000001$ \\
$M\leftarrow \textsf{pmin}(X, K, 4)$ & $=$ & $10000000\,\,\,\,\,00000000$ \\
\hline
\end{tabular}
\end{scriptsize}
\end{center}
\caption{Example of the algorithm and of the \textsf{fnf} operation for $\sigma=4$, $P=01\,10\,01$, $T=01\,10\,11\,11\,10\,11\,01\,10$, $w=16$ and $k=1$. The word $A$ encodes two pattern copies while the word $B_0$ encodes the text substrings $T[0\,\ldots\,2]$ and $T[4\,\ldots\,6]$. Padding fields are underlined. The pattern matches the first and second substring with $1$ and $2$ mismatches, respectively. Since $k=1$, only the first field is nonzero in the \textsf{pmin} output word.}
\label{fig:example}
\end{figure}
Then, we use the \textsf{bsa} operation to count the number of mismatches for each substring, i.e., we
compute a \fword{f} such that each field of
$\bar{m}\log\sigma$ bits contains the number of bits set
(mismatches) in the corresponding block of $\bar{m}$ fields of $A'$.
Observe that in this setting \textsf{bsa} has $O(\log\min(m,\log w /
\log\sigma))$ time complexity. Then, to find all the occurrences with
at most $k$ mismatches we use the \textsf{pmin} operation with the
word $K$ to identify the blocks with a bit count less than or equal to
$k$. Finally, to iterate over all the occurrences we use the
well-known bitwise operation that computes the position of the highest
bit set in a word. Observe that this operation is in $AC^0$ and takes
constant time~\cite{DBLP:journals/tcs/AnderssonMT99}. Hence, our
algorithm has \wrbound time complexity, and it obtains the
$O(\frac{n}{\floor{w/(m\log\sigma)}})$ bound, corresponding to no
overhead for the bitwise operations, for $\log\sigma =
\Omega(\log w)$ or constant $m$.
An example of the algorithm is depicted in Figure~\ref{fig:example}.

We now present the algorithm in the $AC^0$ model. Let $\bar{k}$ be the
smallest power of two greater than $k$. We distinguish two cases: if
$\log\sigma < \log \bar{k} + 1$ we simply run the word-RAM solution. In
$AC^0$ \textsf{bsa} has $O(\log m)$ time complexity and so the
algorithm runs in $O(\frac{n}{\floor{w/(m\log\sigma)}} \log m)$ time.
Otherwise, the algorithm performs the following main steps, for each
$0\le i < n / \ell$:
\begin{enumerate}
\item $\word{X}\leftarrow \word{A} \uxor \word{B}_i$
\item $\word{A'}\leftarrow \textsf{fnf}(\word{X}, \log\sigma)$
\item $\word{H}\leftarrow (\word{H}<< \fsize) \uor \word{A'}$
\item \textbf{if} $i > 0$ and $i\bmod \floor{\log\sigma / \fsize} = 0$
\item \qquad $M\leftarrow \textsf{pmin}(\textsf{ibsa}(\word{H}, \fsize, \bar{m}), K, \fsize)$
\item \qquad $\textbf{report}(M)$
\item \qquad $\word{H}\leftarrow 0$
\end{enumerate}
where in this case $\fsize = \log\bar{k} + 1$ and $\word{H}$ is a word initialized to $0$.
The main difference in this algorithm is that we
report the occurrences every $\floor{\log\sigma / \fsize}$ iterations,
so as to reduce the overhead due to counting the number of mismatches when $\log k = o(\log\sigma)$. To this end, we
compact the fields in the word $\textsf{fnf}(\word{A}\uxor\word{B_i}, \log\sigma)$
into fields of size $\fsize$ in the word $\word{H}$. If $i > 0$ and
$i\bmod \floor{\log\sigma / \fsize} = 0$, i.e., every
$\ell\floor{\log\sigma / \fsize}$ processed substrings, we report the
occurrences as follows. First, observe that the word $H$ contains
$\ell\bar{m}\floor{\log\sigma / \fsize}$ fields of $\fsize$ bits,
encoding the mismatches for the substrings of $T$ of length $m$
corresponding to the words $\word{B}_{i-j}$, for
$j=0,\ldots,\floor{\log\sigma / \fsize}-1$. More precisely, the $l$-th
sequence of $\textsf{fnf}(\word{A} \uxor \word{B}_{i-j}, \log\sigma)$ spans the
fields
$
s, s + \floor{\log\sigma / \fsize}, \ldots, s + \floor{\log\sigma / \fsize}(\bar{m} - 1)\,,
$
where $s = j + (l-1)\bar{m}\floor{\log \sigma / \fsize}$, for
$l=1,\ldots,\ell$. Using a suitable algorithm, i.e, the \textsf{ibsa}
operation, we compute a word such that the last field of each sequence
has value equal to the number of bits set (mismatches) in all the
fields of the sequence if the number of mismatches is less than
$\bar{k}$ and to a value $\ge \bar{k}$ otherwise. Then, we proceed as in the word-RAM algorithm.
We assumed for simplicity that $\fsize$ divides $\log\sigma$ so that
$H$ is a \fword{\fsize}. In general, we have $\log\sigma\bmod f$
unused bits every $\floor{\log\sigma / f}$ fields in $H$. The
algorithm works correctly also in this case, by suitably honoring this
layout in $\word{K}$, \textsf{ibsa} and \textsf{pmin}.
The time complexity of this algorithm is \acbound. It obtains the
$O(\frac{n}{\floor{w/(m\log\sigma)}})$ bound if $\log \min(k,\sigma) \log m =
O(\log\sigma)$.

Finally, we give a variant useful for two extreme cases:
either $k$ or $m-k$ is very small.
More precisely, it is competitive when $k =
o(\log \min(k,\sigma) \log m / \log\sigma)$ or $m - k = o(\log \min(k,\sigma) \log m / \log\sigma)$.
It uses only $AC^0$ instructions.
In this variant, first presented for the case of small $k$, we compute $A$ and $B_i$ using $\bar{m} = m + 1$.
Each block in $A$ and $B_i$ has
thus one padding field and $p = (m+1)\log\sigma$ associated bits. The most significant bit
of the padding field is a sentinel that will signal that there
are more than $k$ mismatches, as
will be shown shortly.
The idea is to parallelize the well-known
sideways addition implementation in which the least significant bit
set is cleared in a
loop\footnote{\url{http://graphics.stanford.edu/~seander/bithacks.html#CountBitsSetKernighan}}.
To this end, we perform the following procedure:
\begin{enumerate}
\item $\word{X}\leftarrow \word{A} \uxor \word{B}_i$
\item $\word{A'}\leftarrow \textsf{fnf}(\word{X}, \log\sigma)$
\item \textbf{for} $i\leftarrow 1$ \textbf{to} $k+1$
\item \qquad $\word{A'} \leftarrow \word{A'} \uor \word{V_p}$
\item \qquad $\word{A'} \leftarrow \word{A'} \uand (\word{A'} - (\word{V_p} >> (p - 1)))$
\item $\word{M} \leftarrow (\word{A'}\uand \word{V_p}) \uxor \word{V_p}$
\item $\textbf{report}(M)$
\end{enumerate}
At each iteration of the loop we add the value $2^{p-1}$
(corresponding to the sentinel bit) to each block in $A'$ and clear
the least significant bit set. In this way, after $k+1$ iterations,
the sentinel bit of any block is set iff the number of mismatches
is at least $k+1$. We then replace the value of each block with $2^{p-1}$
if the sentinel bit is not set and with $0$ otherwise.
The complexity of the described operation is $O(k)$. The time complexity
of this algorithm is $O(\frac{n}{\floor{w/(m\log\sigma)}}\;k)$.
A twin solution handles the case of small $m - k$. The idea is to
find the blocks where the number of matching symbols is at least $m - k$,
which basically consists in using the same method on the bitwise complement of the top bits of
$\word{A'}$.

\section{Applications}

\noindent
The presented technique can be used for several other string matching problems.
We show how to adapt it for particular models in the following subsections.

\subsection{Matching with $k$-mismatches and wildcards}
Assume that the integer alphabet $\Sigma$, of size $\sigma$, contains a
wildcard symbol $\phi$ , i.e., a special symbol that matches any other
symbol of the alphabet. We consider the $k$-mismatches problem with
wildcards~\cite{DBLP:conf/soda/CliffordEPR09}, which consists in reporting all the positions $j$ such that
$|\{0 \leq h < m\ :\ T[j + h]\neq P[h]\wedge T[j + h]\neq \phi\wedge
P[h]\neq \phi\}|\le k$.
Let $A$ and $B_i$ be defined as in Sect.~\ref{sec:the_algorithm}.
The idea is to modify our algorithm so as to reset to zero all the
fields $j$ in $\textsf{fnf}(A \uxor B_i, \log\sigma)$ such that $A[j] = \phi$ or
$B_i[j] = \phi$, since there can be no mismatch in a position where
either a pattern or text wildcard occurs.

In the preprocessing we create two \fword{\log\sigma}s $W_P$ and $H_T$.
A field $W_P[i]$ in $W_P$ is equal to $0$ if $i > \ell\bar{m}$ or
$A[i] = \phi$, to $2^{\log\sigma-1}$ otherwise. A field
$H_T[i]$ in $H_T$ is equal to $\phi$ if $i\le \ell\bar{m}$,
to $0$ otherwise.

At each iteration $i$ of the searching phase, we compute the word $W_T =
\textsf{fnf}(B_i\uxor H_T, \log\sigma)$.
Analogously to $W_P$, a field $W_T[j]$ in $W_T$ is equal to $0$ if
$i > \ell\bar{m}$ or $\word{B}_i[j] = \phi$, to $2^{\log\sigma-1}$ otherwise.

Then, we \band the result of operation $2$ of the algorithm with
$W_P\uand W_T$ (i.e., $\word{A'}\leftarrow \word{A'} \uand (W_P\uand
W_T))$. The rest of
the procedure is unchanged. The overall time complexity is also
unchanged.

\subsection{$\delta$-matching with $k$-mismatches and $(\delta,\gamma)$-matching}

We consider the problem of $\delta$-matching~\cite{CCIMPijcm02,DBLP:conf/cpm/CliffordCI05} with $k$-mismatches.
In this problem we want to report, given an integer $\delta$, all the positions $j$ such
that $|\{0 \leq i < m\ :\ |T[j + i] - P[i]| > \delta\}|\le k$.
In the exact case, i.e., when $k=0$, this is equivalent to matching under the $L_\infty$ distance~\cite{DBLP:conf/cpm/CliffordCI05}.
In $\delta$-matching any two characters $t$ and $p$ are defined to match iff $|t-p| \leq \delta$.
Note that in the algorithm to be presented in this section we can allow $\delta_i$ to be different for each pattern position $i$,
while usually in $\delta$-matching the allowed error $\delta$ is the same for each character.
This also yields an alternate solution to matching with wildcards in the pattern, by simply
using $\delta_i = \sigma-1$ for pattern positions $i$ corresponding to wildcards, and $\delta_i = 0$ elsewhere.

The idea is to compute the absolute difference $|A[j] - B_i[j]|$ for
each field $j$ using \textsf{pvmax} and \textsf{pvmin},
i.e., we compute a \fword{\log\sigma} $X$ such that $X[j]=\max(A[j], B_i[j]) - \min(A[j], B_i[j])$. Then we
replace each difference with $2^{\log\sigma -1}$ if it is greater than
$\delta_{j\mod m}$ and with $0$ otherwise, using \textsf{pmin} and a
\bxor operation. In this way we can count the number of symbols that do not $\delta$-match using \textsf{ibsa} or \textsf{bsa}.
In the preprocessing phase we compute $D'[j] = \delta_j$ and construct
its packed representation $\word{D}$, a \fword{\log\sigma} holding $\ell$ copies of $D'$.
Let $\word{W}$ be a \fword{\log\sigma} such that $W[i]$ is equal to
$2^{\log\sigma-1}$ if $i\le \ell m$, to $0$ otherwise.
Then at iteration $i$ of the searching phase we compute
\begin{enumerate}
\item $\word{X} \leftarrow
\textsf{pvmax}(\word{A},\word{B_i},\log\sigma) -
\textsf{pvmin}(\word{A},\word{B_i},\log\sigma)$
\item $\word{A'} \leftarrow \textsf{pmin}(\word{X}, \word{D},\log\sigma) \uxor \word{W}$
\end{enumerate}
The result is that each field of $\word{A'}$ is equal to $2^{\log\sigma-1}$ iff the corresponding
pattern and text characters do not $\delta$-match.
The rest of the algorithm is as before, only the steps 1--2 of either
of the main algorithms (for $AC^0$ and word-RAM models) are replaced
with the two steps above. The time complexities remain the same.

If we are interested in the (more conventional) exact $\delta$-matching variant
(i.e.\ assume that $k=0$), we can improve
the time to $O(\frac{n}{\floor{w/(m/\log\sigma)}})$ using the following algorithm:
\begin{enumerate}
\item $\word{X} \leftarrow
\textsf{pvmax}(\word{A},\word{B_i},\log\sigma) -
\textsf{pvmin}(\word{A},\word{B_i},\log\sigma)$
\item $\word{A'} \leftarrow \textsf{pmin}(\word{X}, \word{D},\log\sigma) \uxor \word{W}$
\item $M\leftarrow \textsf{fnf}(\word{A'}, m\log\sigma) \uxor \word{V_{m\log\sigma}}$
\item $\textbf{report}(M)$
\end{enumerate}

The \textsf{fnf} operation interprets the word $\word{A'}$ as a
\fword{m\log\sigma}, and returns a \fword{m\log\sigma} where
the $i$-th field is equal to $2^{m\log\sigma-1}$ if at least one character of
the $i$-th pattern copy did not $\delta$-match, and to $0$ otherwise.
The \bxor operation then inverts the fields' values, so that a field is
equal to $2^{m\log\sigma-1}$ if all the characters of the
corresponding pattern copy did $\delta$-match.

We note that a close relative to $\delta$-matching is less-than matching, where characters
$p$ and $t$ match if $p \leq t$.
This model has applications in other pattern matching problems, see e.g.\
\cite{DBLP:conf/focs/AmirALLL97}.
The less-than matching problem can be easily solved with our methods in the same way as
$\delta$-matching, that is,
the first two lines are simply replaced with
$A' \leftarrow \textsf{pmin}(A, B_i, \log\sigma) \uxor \word{W}$. It is also possible to solve
$\delta$-matching, with the same time complexity, by combining less-than and greater-than matching.

The $\gamma$-matching problem consists in, given an integer
$\gamma$, finding all the positions $j$ such that $\sum_{0 \leq i < m}
|T[j + i] - P[i]| \le \gamma$. If both $\delta$ and $\gamma$
conditions must hold, we speak of $(\delta,\gamma)$-matching. There
are many algorithms devoted to this model, see
e.g.\ \cite{CINPSjda04}. In order to solve $\gamma$-matching
we need to sum the fields of each pattern copy in $\word{X}$ and
compare each value against $\gamma$ (note that we need to check also
the $\delta$ condition). If we do not use field compaction and
deferred reporting of occurrences, this is easiest to do in the same
way as the first widening phase of \textsf{bsa} operation, i.e.\ by
simply shifting and adding the fields in parallel in
$O(\log m)$
time, giving
$O(\frac{n}{\floor{w/(m\log\sigma)}} \log m)$
total time in $AC^0$.
This result can be improved in both $AC^0$ and word-RAM models.
We first define one more operation:

\medskip
\noindent \emph{interleave two words}
(\textsf{interleave}$(A,B,Z,f)$): given three \fword{f}s $\word{A}$,
$\word{B}$ and $\word{Z}$, return a \fword{f} $W$ such that $W[i]$ is
equal to $A[i]$ if $Z[i] = 0$, and to $B[i]$ otherwise. This can be
implemented in $O(1)$ time as follows:
\begin{enumerate}
\item $\word{Z} \leftarrow \textsf{fnf}(\word{Z}, f)$
\item $\word{Z} \leftarrow (\word{Z} - (\word{Z} >> (f-1))) \uor \word{Z}$
\item $\word{W} \leftarrow (\word{A} \uand \unot \word{Z}) \uor (\word{B} \uand \word{Z})$
\end{enumerate}
The idea is to first find all the pattern copies with at least one
difference greater than $\delta$, by computing the
\fword{\bar{m}\log\sigma} $Z = \textsf{fnf}(\word{A'},
\bar{m}\log\sigma)$, as in the algorithm described before for
$\delta$-matching. Then, we prevent all these pattern copies from
$\gamma$-matching by replacing all the corresponding differences in
$X$ with $\delta$. The effect is that the sum of the fields for a
pattern copy with at least one difference greater than $\delta$ is
equal to $m \delta$ instead of the exact sum. This works because in
$\gamma$-matching it always holds that $m \delta > \gamma$, as
otherwise the $\gamma$ condition does not prune anything. To this end,
using $Z$ and \textsf{interleave}, we interleave the words $X$ and
$D$, interpreted as \fword{m\log\sigma}s, into the \fword{\log\sigma}
$X'$. In this way, a field $X'[i]$ is equal to $X[i]$ if no difference
is greater than $\delta$ for the corresponding pattern copy (i.e., if
$Z[\floor{i / \bar{m}}] = 0$), and to $D[i]$ otherwise.

In word-RAM we then use \textsf{bsa} and \textsf{pmin} to accumulate
the absolute differences in $X'$ and compare the sums against the
threshold value $\gamma$ in parallel. Note that we need to adjust
\textsf{bsa} to use as $r$ the smallest power of two greater than or
equal to $\log(w \delta+1)/\log\sigma$, so as to not cause overflows.
The algorithm is
\begin{enumerate}
\item $\word{X} \leftarrow \textsf{pvmax}(\word{A},\word{B_i},\log\sigma) -
\textsf{pvmin}(\word{A},\word{B_i},\log\sigma)$
\item $\word{A'} \leftarrow \textsf{pmin}(\word{X}, \word{D},\log\sigma) \uxor \word{W}$
\item $\word{Z} \leftarrow \textsf{fnf}(\word{A'}, \bar{m}\log\sigma)$
\item $\word{X'} \leftarrow \textsf{interleave}(\word{X}, \word{D}, \word{Z}, \bar{m}\log\sigma)$
\item $M\leftarrow \textsf{pmin}(\textsf{bsa}(\word{X'}, \log\sigma, \bar{m}), G, \fsize)$
\item $\textbf{report}(M)$
\end{enumerate}
where $\fsize = \bar{m}\log\sigma$ and $\word{G}$ is a \fword{\fsize}
containing a copy of the integer $\gamma$ in each field.
The time complexity is
$O(\frac{n}{\floor{w/(m\log\sigma)}} \log \min(m, \log(w\delta)/\log\sigma))$ which again obtains
$O(\frac{n}{\floor{w/(m\log\sigma)}})$ time for $\log\sigma = \Omega(\log(w\delta))$ or constant $m$.

\medskip

Consider now the $AC^0$ model. We use an approach similar to the one
used in the $k$-mismatches case, the only difference is that we need
more bits to represent the accumulated sums. That is, we replace
$\bar{k}$ with $\bar{\gamma} = 2^{\ceil{\log(\gamma+1)}}$ (the
smallest power of two greater than $\gamma$), and use $\fsize =
\log\bar{\gamma}+1$. The \textsf{ibsa} operation then works correctly,
i.e.\ accumulates the absolute differences without overflowing the
sums, provided that all characters $\delta$-match, as otherwise the
corresponding absolute difference may be $\sigma-1$, which in turn can
be larger than $\bar{\gamma}$.
This holds because no difference is greater than $\delta$ in $X'$.
The complete pseudocode follows.
\begin{enumerate}
\item $\word{X} \leftarrow \textsf{pvmax}(\word{A},\word{B_i},\log\sigma) -
\textsf{pvmin}(\word{A},\word{B_i},\log\sigma)$
\item $\word{A'} \leftarrow \textsf{pmin}(\word{X}, \word{D},\log\sigma) \uxor \word{W}$
\item $\word{Z} \leftarrow \textsf{fnf}(\word{A'}, \bar{m}\log\sigma)$
\item $\word{X'} \leftarrow \textsf{interleave}(\word{X}, \word{D}, \word{Z}, \bar{m}\log\sigma)$
\item $\word{H}\leftarrow (\word{H} << \fsize) \uor \word{X'}$
\item \textbf{if} $i > 0$ and $i\bmod \floor{\log\sigma / \fsize} = 0$
\item \qquad $M\leftarrow \textsf{pmin}(\textsf{ibsa}(\word{H}, \fsize, \bar{m}), G, \fsize)$
\item \qquad $\textbf{report}(M)$
\item \qquad $\word{H}\leftarrow 0$
\end{enumerate}
The total time is $O(\frac{n}{\floor{w/(m\log\sigma)}} (1+\log m \log\gamma / \log\sigma))$,
This becomes $O(\frac{n}{\floor{w/(m\log\sigma)}})$ for $\log m\log\gamma = O(\log\sigma)$.

We can also combine the two models, $\delta$-matching with
$k$-mismatches and $(\delta,\gamma)$-matching, to obtain
$(\delta,k,\gamma)$-matching. In this model we limit the number of
characters not $\delta$-matching by $k$, and the accumulated sum of
the absolute differences by $\gamma$. The basic idea is to compute two
match vectors, $M_\delta$ and $M_\gamma$, take their bitwise \band as
$M \leftarrow M_\delta \uand M_\gamma$ and then report the occurrences
with respect to $M$. The vector $M_\delta$ can be computed as was
already shown. To compute $M_\gamma$ we just skip the
\textsf{interleave} operation to take the absolute differences
raw without saturating them with $\delta$. In $AC^0$ we can use the
basic algorithm to compute $M_\gamma$ in
$O(\frac{n}{\floor{w/(m\log\sigma)}} \log m)$ time, which dominates
the total time. However, since the sums need more bits now, there is a
significant overhead in the case of the word-RAM algorithm and of the
improved $AC^0$ algorithm. In particular, in the case of the word-RAM
algorithm, the time complexity becomes
$O(\frac{n}{\floor{w/(m\log\sigma)}} \log \min(m,
\log(w\sigma)/\log\sigma))$, i.e., slightly worse. Instead, in the
case of the improved $AC^0$ algorithm, the overhead makes the
algorithm useless. However, we can still manage to obtain
$O(\frac{n}{\floor{w/(m\log\sigma)}} (1+\log m \log\gamma /
\log\sigma))$ time by modifying the model so that we accumulate the
absolute differences only on $\delta$-matching character positions.
This can be easily done with the tools already presented, namely using
the \textsf{interleave} operation to set non-$\delta$-matching
character positions to 0 in word $\word{X'}$.



\section{Conclusion}
\noindent
We presented a novel technique for approximate pattern matching with
$k$-mismatches when the text is given in packed form. Assuming the
pattern is short enough, it is possible to achieve a sublinear search
time, if several pattern copies are matched against different text
substrings at the same time. We described variants of our simple
method in the $AC^0$ and word-RAM models and also considered the case
when the number $k$ of allowed errors is small. Moreover, we showed
how to adapt our algorithms to other matching models, including
approximate matching with wildcard (don't-care) symbols,
$\delta$-matching with $k$-mismatches and $(\delta, \gamma)$-matching.

\section{Acknowledgments}

We thank two anonymous reviewers and Djamal Belazzougui for helpful comments.

\bibliographystyle{abbrv}
\bibliography{hamming}

\begin{thebibliography}{10}

\bibitem{DBLP:conf/focs/AmirALLL97}
A.~Amir, Y.~Aumann, G.~M. Landau, M.~Lewenstein, and N.~Lewenstein.
\newblock Pattern matching with swaps.
\newblock In {\em FOCS}, pages 144--153. IEEE Computer Society, 1997.

\bibitem{DBLP:journals/jal/AmirLP04}
A.~Amir, M.~Lewenstein, and E.~Porat.
\newblock Faster algorithms for string matching with \emph{k} mismatches.
\newblock {\em J. Algorithms}, 50(2):257--275, 2004.

\bibitem{DBLP:journals/tcs/AnderssonMT99}
A.~Andersson, P.~B. Miltersen, and M.~Thorup.
\newblock Fusion trees can be implemented with {AC}$^{\mbox{0}}$ instructions
  only.
\newblock {\em Theor. Comput. Sci.}, 215(1-2):337--344, 1999.

\bibitem{DBLP:journals/cacm/Baeza-YatesG92}
R.~A. Baeza-Yates and G.~H. Gonnet.
\newblock A new approach to text searching.
\newblock {\em Commun. ACM}, 35(10):74--82, 1992.

\bibitem{Belazzougui2012jda}
D.~Belazzougui.
\newblock Worst-case efficient single and multiple string matching on packed
  texts in the word-{RAM} model.
\newblock {\em J. Discrete Algorithms}, 14:91--106, 2012.

\bibitem{Ben-KikiBBGGW2011}
O.~Ben-Kiki, P.~Bille, D.~Breslauer, L.~Gasieniec, R.~Grossi, and O.~Weimann.
\newblock Optimal packed string matching.
\newblock In S.~Chakraborty and A.~Kumar, editors, {\em FSTTCS}, volume~13 of
  {\em LIPIcs}, pages 423--432. Schloss Dagstuhl - Leibniz-Zentrum fuer
  Informatik, 2011.

\bibitem{Bille2011}
P.~Bille.
\newblock Fast searching in packed strings.
\newblock {\em J. Discrete Algorithms}, 9(1):49--56, 2011.

\bibitem{BreslauerGG2012}
D.~Breslauer, L.~Gasieniec, and R.~Grossi.
\newblock Constant-time word-size string matching.
\newblock In J.~K{\"a}rkk{\"a}inen and J.~Stoye, editors, {\em CPM}, volume
  7354 of {\em Lecture Notes in Computer Science}, pages 83--96. Springer,
  2012.

\bibitem{CCIMPijcm02}
E.~Cambouropoulos, M.~Crochemore, C.~S. Iliopoulos, L.~Mouchard, and Y.~Pinzon.
\newblock Algorithms for computing approximate repetitions in musical
  sequences.
\newblock {\em International Journal of Computer Mathematics},
  79(11):1135--1148, 2002.

\bibitem{DBLP:conf/cpm/CliffordCI05}
P.~Clifford, R.~Clifford, and C.~S. Iliopoulos.
\newblock Faster algorithms for delta, gamma-matching and related problems.
\newblock In A.~Apostolico, M.~Crochemore, and K.~Park, editors, {\em CPM},
  volume 3537 of {\em Lecture Notes in Computer Science}, pages 68--78.
  Springer, 2005.

\bibitem{DBLP:conf/soda/CliffordEPR09}
R.~Clifford, K.~Efremenko, E.~Porat, and A.~Rothschild.
\newblock From coding theory to efficient pattern matching.
\newblock In C.~Mathieu, editor, {\em SODA}, pages 778--784. SIAM, 2009.

\bibitem{CINPSjda04}
M.~Crochemore, C.~S. Iliopoulos, G.~Navarro, Y.~J. Pinzon, and A.~Salinger.
\newblock Bit-parallel $(\delta,\gamma)$-matching and suffix automata.
\newblock {\em J. Discrete Algorithms}, 3(2-4):198--214, 2005.

\bibitem{Fredriksson2002}
K.~Fredriksson.
\newblock Faster string matching with super-alphabets.
\newblock In A.~H.~F. Laender and A.~L. Oliveira, editors, {\em SPIRE}, volume
  2476 of {\em Lecture Notes in Computer Science}, pages 44--57. Springer,
  2002.

\bibitem{DBLP:journals/ejc/FredrikssonG13}
K.~Fredriksson and S.~Grabowski.
\newblock Exploiting word-level parallelism for fast convolutions and their
  applications in approximate string matching.
\newblock {\em Eur. J. Comb.}, 34(1):38--51, 2013.

\bibitem{DBLP:journals/ipl/GrabowskiF08}
S.~Grabowski and K.~Fredriksson.
\newblock Bit-parallel string matching under {H}amming distance in
  ${O}(n\ceil{m/w})$ worst case time.
\newblock {\em Inf. Process. Lett.}, 105(5):182--187, 2008.

\bibitem{HS1986}
W.~D. Hillis and G.~L. Steele, Jr.
\newblock Data parallel algorithms.
\newblock {\em Commun. ACM}, 29(12):1170--1183, 1986.

\bibitem{DBLP:journals/jea/HyyroFN05}
H.~Hyyr{\"o}, K.~Fredriksson, and G.~Navarro.
\newblock Increased bit-parallelism for approximate and multiple string
  matching.
\newblock {\em ACM Journal of Experimental Algorithmics}, 10(2.6):1--27, 2005.

\bibitem{PS1980}
W.~Paul and J.~Simon.
\newblock Decision trees and random access machines.
\newblock In {\em ZUERICH: Proc. Symp. Logik und Algorithmik}, pages 331--340,
  1980.

\bibitem{DBLP:conf/wea/Vigna08}
S.~Vigna.
\newblock Broadword implementation of rank/select queries.
\newblock In C.~C. McGeoch, editor, {\em WEA}, volume 5038 of {\em Lecture
  Notes in Computer Science}, pages 154--168. Springer, 2008.

\end{thebibliography}

\end{document}